\documentclass[11pt]{article}
\usepackage[margin=1in]{geometry}
\usepackage[T1]{fontenc}
\usepackage{lmodern}
\usepackage{amsmath,amssymb,amsthm,mathtools}
\usepackage{microtype}
\usepackage{enumitem}
\usepackage{booktabs}
\usepackage{array}
\usepackage{natbib}
\usepackage[colorlinks=true,linkcolor=blue,citecolor=blue,urlcolor=blue]{hyperref}
\hypersetup{pdftitle={The Complexity of Quadratic KKT Points with Bounded Integer Coefficients},pdfauthor={}}
\setlist[itemize]{leftmargin=1.4em,itemsep=3pt,topsep=5pt}
\setlist[enumerate]{leftmargin=1.7em,itemsep=3pt,topsep=5pt}
\newtheorem{theorem}{Theorem}[section]
\newtheorem{lemma}[theorem]{Lemma}
\newtheorem{proposition}[theorem]{Proposition}

\theoremstyle{definition}
\newtheorem{definition}[theorem]{Definition}
\theoremstyle{remark}

\newcommand{\CLS}{\mathsf{CLS}}
\newcommand{\PPAD}{\mathsf{PPAD}}
\newcommand{\TFNP}{\mathsf{TFNP}}
\newcommand{\PLS}{\mathsf{PLS}}

\newcommand{\NP}{\mathsf{NP}}

\newcommand{\xbf}{\mathbf{x}}
\newcommand{\cbf}{\mathbf{c}}

\newcommand{\qkkt}{{\normalfont\scshape Quadratic-KKT}}

\usepackage[most]{tcolorbox}

\usepackage{todonotes}

\usepackage{tikz}
\usetikzlibrary{positioning,decorations.pathreplacing}

\title{The Complexity of KKT Points of Quadratic Programs with Polynomially-Bounded Coefficients}

\author{Argyrios Deligkas, Michail Fasoulakis, Stavros D. Ioannidis\\
 Royal Holloway, University of London, United Kingdom}
\date{}
\begin{document}
\maketitle
\begin{abstract}
Computing a Karush–Kuhn–Tucker (KKT) point of a quadratic polynomial over the unit hypercube is known to be \(\CLS\)-complete, when the coefficients of the polynomial are encoded in binary. We prove that the problem remains \(\CLS\)-complete even when the coefficients are encoded in unary.
This is the first problem proven to remain $\CLS$-hard when the input is given in unary.
\end{abstract}

\section{Introduction}
Quadratic programming is possibly one of the most fundamental problems in continuous optimization, with applications ranging from game theory and machine learning, to engineering and computational chemistry~\cite{DBLP:journals/nature/GusevADACKPDDSR23}.
One of the most basic forms of the problem is defined via ``{\em box constraints}''~\cite{andjouh2022adaptive,burer2009nonconvex}.
There, the problem is defined via an $n \times n$ symmetric coefficient-matrix $Q$, and an $n$-dimensional coefficient-vector $\cbf$. The goal is to find a point in the $n$-dimensional hypercube that minimizes the corresponding quadratic program, formally:
\begin{equation}
\label{eq:QP_formulation}
    \min_{\xbf \in [0,1]^n} P(\xbf):= \xbf^T Q \xbf + \cbf^T\xbf.
\end{equation}

Since the complexity of the problem was proven to be $\NP$-hard more than fifty years ago~\cite{sahni1972some,sahni1974computationally}, the relaxed solution of {\em Karush-Kuhn-Tucker}, a.k.a., KKT, points was considered.
There, instead of a global optimum of $P(\xbf)$, the solution can be any {\em first-order stationary point}, a.k.a. fixed point of gradient descent, of the polynomial.
In a recent celebrated result, \cite{FGHS25} showed that computing a  KKT point of~\eqref{eq:QP_formulation}, even in the approximate sense, is $\CLS$-complete.
This serves as an indication that it is unlikely the problem to have a polynomial-time algorithm.

The complexity class $\CLS$, standing for Continuous Local Search, was introduced by Daskalakis and Papadimitriou~\cite{DP11} to capture the complexity of search problems that belong to both $\PPAD$~\cite{papadimitriou1994complexity} and $\PLS$~\cite{johnson1988easy} yet no polynomial-time algorithm is known.
In other words, the existence of the solution is guaranteed either via a fixed point argument {\em or} via a potential function argument.
In their breakthrough result, \cite{FGHS25} proved that $\CLS = \PPAD \cap \PLS$. 

Since then, a series of works established $\CLS$-hardness for  congestion games~\cite{babichenko2021settling} and equilibrium computation problems~\cite{elkind2022simultaneous,tewolde2023computational}, and very recently, in yet another breakthrough result, for KKT points in {\em bilinear} polynomials, i.e., polynomials without any square term~\cite{APY26-net_coordination}. 
However, in all of the above-mentioned results the input of the problem was given with {\em binary encoding}; hence the input could include coefficients of exponential precision. 
In fact, the correctness of the reduction of~\cite{FGHS25} heavily relies on the existence of exponential ratios between the coefficients and it explicitly asks as an open problem whether $\CLS$-hardness still holds when the input is given in unary.
Additionally, for the case of KKT points for bilinear polynomials, binary encoding is actually {\em required} in order to establish $\CLS$-hardness~\cite{APY26-net_coordination}.
If the input is given in unary the bilinear polynomials become polynomial-time solvable.

To see the latter, observe that a bilinear polynomial is affine in each coordinate separately. So it always has a global minimum, and hence a KKT point, at a Boolean vector, i.e.,  a solution where $x_i \in \{0,1\}$ . Moreover, a Boolean point is a KKT point precisely when no single bit flip decreases the objective. If the coefficients are integers,  each improving flip decreases the objective by at least one. With polynomially bounded coefficients, the range of the objective on Boolean points is also polynomially bounded, and this elementary local search procedure terminates in polynomial time.

Hence, the open question from~\cite{FGHS25} becomes even more intriguing: {\em is binary encoding required to establish $\CLS$-hardness in general, or can we prove hardness even when the input is given in unary?}

\paragraph{\bf Our result and its implications.}
We answer the open question of \cite{FGHS25} and prove that finding a KKT point of~\eqref{eq:QP_formulation} remains $\CLS$-hard even when all coefficients, of $Q$ and $\cbf$, are given in unary. 
This is the first problem where the input is given in unary and it is $\CLS$-hard.
As a consequence, our result rules out the existence of pseudopolynomial-time algorithm for KKT points. Our result provides an interesting dichotomy: allowing square terms produces a sharp change in complexity of KKT points. When the input is given in unary, bilinear KKT admits a polynomial time algorithm, whereas quadratic KKT remains $\CLS$-complete; see Table~\ref{tab:classification}.
Thus, our result implies that the reduction of~\cite{APY26-net_coordination} is essentially ``optimal'': the exponential blow-up in the coefficients of the produced bilinear polynomial is {\em necessary}. 
Finally, our result  can be seen as a complement to the recent result of~\cite{APY26-net_coordination}.
The equality $\CLS=\PPAD\cap\PLS$ gives two possible approaches to solving problems in $\CLS$: one through fixed point computation and another through local search. A restriction that makes either approach efficient is enough to obtain a polynomial-time algorithm; 
for the latter, \cite{DBLP:journals/siamcomp/SchafferY91,10.1145/100216.100274} observed that many problems become tractable under unary encoding. 
Along with~\cite{APY26-net_coordination}, we establish that hardness for KKT points persists when either square terms or large integer coefficients are excluded separately, but excluding both makes the problem tractable.

\begin{table}[t]
\centering
\small
\begin{tabular}{@{}>{\raggedright\arraybackslash}p{0.24\textwidth}>{\raggedright\arraybackslash}p{0.32\textwidth}>{\raggedright\arraybackslash}p{0.34\textwidth}@{}}
\toprule
Objective & Binary-encoded coefficients & Unary-encoded coefficients \\
\midrule
Quadratic & $\CLS$-complete \cite{FGHS25} & $\CLS$-complete; \textbf{this work} \\ [3pt]
Bilinear & $\CLS$-complete \cite{APY26-net_coordination} & Polynomial time; Boolean solution \\
\bottomrule
\end{tabular}
\caption{The complexity of computing exact KKT points over the unit cube.}
\label{tab:classification}
\end{table}

\section{Technical overview}
Our starting point is the $\CLS$-complete problem \qkkt problem from~\cite{FGHS25}, where we are given a quadratic polynomial $P$ with $n$ variables and binary-encoded coefficients, i.e., coefficients can have exponential precision. 
We construct a quadratic polynomial that {\em preserves all stationary points} of $P$ and has polynomially-bounded integer coefficients. 
The high-level idea is to simulate the exponentially-precise coefficients of $P$ with polynomially-many auxiliary variables: we ``push'' the precision of the coefficients to the values of the auxiliary variables at any KKT point of the constructed polynomial. 
In other words, at any KKT point of the constructed polynomial, the auxiliary variables need exponential precision and thus they can simulate the coefficients or the original instance.

More concretely now, our reduction consists of three main steps.

\medskip
\noindent
{\bf Step 1: Normalization.} 
As first step, we normalize the coefficients of the input polynomial $P$ to take values in $(-1,1)$, without changing its KKT points. 
This is done via a ``standard'' way.
This step is crucial for our reduction in order to guarantee the solution space to be the unit hypercube; without normalization the next steps of the reduction would have required exponentially large numbers as solution for the auxiliary variables.

\medskip 
\noindent
{\bf Step 2: Dyadic decomposition.}
The next step is to decompose the normalized polynomial as a sum of monomials with coefficients of the form \(2^{-k}\), where \(k\) is polynomially-bounded by the encoding length of the input. In fact, $k$ is bounded by the maximum number of bits required for any coefficient of $P$. 
This decomposition of $P$ allows us to focus on each monomial independently and introduce independent auxiliary variables. 
Hence, every monomial is of the form \(g(x,y) = 2^{-k}xy\), where $x$ may also be equal to $y$, and its partial derivatives are
\[\frac{\partial g}{\partial x} = 2^{-k}y 
\quad \text{and} \quad \frac{\partial g}{\partial y} = 2^{-k}x.
\]
Observe, each variable contributes to the gradient of \(g\) a value scaled by \(2^{-k}\), i.e., a value obtained by {\em halving the value $k$ times}. 
Our goal then boils down to simulate this halving effect using integer coefficients by introducing a polynomial number of auxiliary variables. 

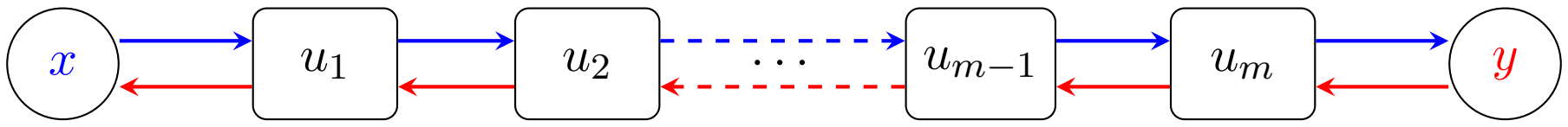
\begin{figure}[h]
\centering
\begin{tikzpicture}[>=stealth,
    endpoint/.style={
        draw,
        circle,
        minimum size=0.85cm
    },
    gate/.style={
        draw,
        rounded corners=3pt,
        minimum width=1.1cm,
        minimum height=0.85cm
    },
    xsignal/.style={->, thick, blue},
    ysignal/.style={->, thick, red}
]

\node[endpoint] (x) at (0,0) {\textcolor{blue}{$x$}};
\node[gate] (u1) at (2,0) {$u_1$};
\node[gate] (u2) at (4,0) {$u_2$};

\node at (5.5,0) {$\cdots$};

\node[gate] (um1) at (7,0) {$u_{m-1}$};
\node[gate] (um) at (9,0) {$u_m$};
\node[endpoint] (y) at (11,0) {\textcolor{red}{$y$}};

\draw[xsignal]
    ([yshift=5pt]x.east) --
    ([yshift=5pt]u1.west);

\draw[xsignal]
    ([yshift=5pt]u1.east) --
    ([yshift=5pt]u2.west);

\draw[xsignal, dashed]
    ([yshift=5pt]u2.east) --
    ([yshift=5pt]um1.west);

\draw[xsignal]
    ([yshift=5pt]um1.east) --
    ([yshift=5pt]um.west);

\draw[xsignal]
    ([yshift=5pt]um.east) --
    ([yshift=5pt]y.west);

\draw[ysignal]
    ([yshift=-5pt]u1.west) --
    ([yshift=-5pt]x.east);

\draw[ysignal]
    ([yshift=-5pt]u2.west) --
    ([yshift=-5pt]u1.east);

\draw[ysignal, dashed]
    ([yshift=-5pt]um1.west) --
    ([yshift=-5pt]u2.east);

\draw[ysignal]
    ([yshift=-5pt]um.west) --
    ([yshift=-5pt]um1.east);

\draw[ysignal]
    ([yshift=-5pt]y.west) --
    ([yshift=-5pt]um.east);
\end{tikzpicture}
\caption{Two-way propagation of values through auxiliary variables.}
\label{fig:chain}
\end{figure}

Before we introduce the next step of the reduction, which simulates the halving effect, let us provide some high-level intuition for it.
We can think of the variables \(x\) and \(y\) of a monomial as two gates that transmit signals to each other, using a {\em chain} of auxiliary variables as shown in Figure~\ref{fig:chain}. 

\noindent
A ``blue'' signal originates from \(x\) and propagates towards \(y\), while getting halved each time it passes through an intermediate gate. Similarly, a ``red'' signal originates from \(y\) and propagates towards \(x\), undergoing the same halving process. The goal is to create a gadget that: a) preserves the KKT points of $P$ and b) does not create any additional KKT points.
The next step of the reduction formally achieves this.

\medskip
\noindent
\textbf{Step 3: The quadratic chain gadget.}
For each monomial \(g(x,y) = s2^{-k}xy\), where \(s\in\{-1,1\}\), we introduce \(m = k+2\) auxiliary variables \(u_1,\ldots u_m\), denoted ${\mathbf u}$ for brevity, each constrained in $[-1,1]$, and construct the following quadratic {\em chain gadget}:
\[
G(x,y,\mathbf u) = 3 \sum_{i=1}^{m-1}(2u_i-u_{i+1})^2+9  u_m^2-6x u_1+6sy u_{m}+x^2+y^2.
\]
Each chain gadget satisfies the following property which is crucial for our purposes: for every fixed values of $x$ and $y$ the function $G$ is {\em strictly convex} with respect to ${\mathbf u}$.
This implies that, for fixed $x,y$, $G$ admits a {\em unique} minimizer, denoted $\mathbf u^*(x,y)$, which is the {\em unique KKT} point of $G$ as well. 
Crucially, the choice of the coefficients of $G$ ensure that $\mathbf u^*(x,y)$ is {\em always} inside $[-1,1]^m$.
In fact, we derive a closed form formula for the auxiliary variables where for $i \in [m]$ we get that 
\[u_i=\frac13\Big(2^{1-i} x-s2^{i-m}y\Big).
\]
Thus, at any KKT point of $G$, the values of the auxiliary variables require exponential precision, which was our initial goal.
The term \(2^{1-i}\) represents the ``blue'' signal from Figure~\ref{fig:chain}, which halves as \(i\) increases from 1 to \(m\), while the term \(2^{i-m}\) represents the ``red'' signal. 
The detailed analysis is presented in Lemma~\ref{lem:interior}. 

As a side note, let us highlight that our chain gadget guarantees more than just equality of the derivatives: it establishes the stronger identity that \(2^{-k}xy\) is equal to \(G(x,y,\mathbf u^*(x,y))\), which can be of independent interest (see Lemma \ref{lem: g equals G with minimizer}).

\medskip
\noindent
{\bf Putting everything together.}
To complete our reduction, the final constructed polynomial is the summation of all chain gadgets of the dyadic monomials. 
The resulting polynomial has integer coefficients of polynomially bounded magnitude. 
To see this, observe the following: a) the coefficients in every chain gadget are constants; b) there are polynomially many chain gadgets; c) the length of each gadget is polynomial.
Furthermore, since the auxiliary variables are disjoint across gadgets, the arguments from Step 3 for KKT points apply independently to them. 

\section{Preliminaries}
\label{sec:preliminaries}

For any positive integer \(n\), we denote \([n] = \{1,\ldots n\}\). 
We consider quadratic polynomials over \(n\) variables \(\mathbf x = (x_i)_{i\in[n]}\) that are given as a sum of monomials: 
\begin{equation}\label{the original polynomial}
P(\mathbf x) = \sum_{i=1}^{n}q_{ii} x_i^2 + \sum_{1\le i<j\le n}q_{ij}x_ix_j + \sum_{i=1}^{n}c_i x_i,  
\end{equation}
where for all $i,j\in[n]$ the coefficients \(q_{ii},q_{ij},c_i\) are assumed to be rational numbers. 

We follow the definition of \emph{box-constrained} KKT point of \(P\) from~\cite{FGHS25}.

\begin{definition}[Box-constrained KKT points]\label{def:KKT}
For a quadratic polynomial \(P\), as expressed in (\ref{the original polynomial}), consider the constrained optimization problem \(\min_{\mathbf x\in [0,1]^n}P(\mathbf x)\). A point \(\mathbf x^*\in[0,1]^n\) is a Karush–Kuhn–Tucker (KKT) point of this optimization problem if for every coordinate \(i\in [n]\) the following conditions are satisfied:
\begin{equation}\label{def: KKT conditions}
\frac{\partial P}{\partial x_i}(\mathbf{x}^*) \geq 0 \text{ if }  x_i^*= 0, \quad \frac{\partial P}{\partial x_i}(\mathbf{x}^*) = 0\text{ if } x_i^* \in(0,1),  \quad \frac{\partial P}{\partial x_i}(\mathbf{x}^*) \leq 0\text{ if } x_i^* = 1.
\end{equation}
\end{definition}

As it was formally explained in~\cite{FGHS25}, under the above mentioned definition a KKT point always does exist.

\paragraph{Binary Vs Unary encoding.}
A quadratic polynomial is represented by an explicit list of its monomials and their corresponding rational coefficients. 
Hence, the {\em input size} is given by the number of variables, which is $n$, plus the size of encoding the coefficients.
When the coefficients are encoded in {\em binary}, i.e., as bitstrings, they can represent numbers of {\em exponential precision}, with respect of the input size. 
For example, the ratio between two coefficients can be \(2^{\Omega(n)}\).
On the other hand, when the coefficients are given in {\em unary}, they have only {\em polynomially-bounded precision}. In other words, $P(\xbf)$ can be represented using integer-coefficients with absolute values bounded by the unary encoding length \(N\) of the input polynomial.
Hence, in this case the ratio between any two coefficients is bounded by $\text{poly}(N)$.

\medskip
We proceed with the definition of \qkkt\, the main computational problem of interest, and the current state of the art for its complexity.

\begin{definition}[\qkkt]\label{def:quadratickkt}
Given a quadratic polynomial \(P\) over \(n\) variables, compute a KKT point of the optimization problem  \(\min_{\mathbf x\in[0,1]^n}P(\mathbf x)\).
\end{definition}

\begin{theorem}[\cite{FGHS25}]
\label{thm:quadratic-kkt-cls}
\qkkt\ is $\CLS$-complete, when the input polynomial is given in binary.
\end{theorem}

\section{The $\CLS$-hardness reduction}

In this section we establish the main theorem of our paper.

\begin{theorem}\label{main theorem}
\qkkt\ is $\CLS$-complete even when the input is given in unary.
\end{theorem}

We present a polynomial-time reduction from \qkkt\ given with binary-encoded coefficients. Specifically, given an instance defined by a quadratic polynomial \(P\), we construct a quadratic polynomial \(\mathcal H\) such that (i) the unary encoding length of \(\mathcal H\) is polynomially bounded in the encoding length of \(P\), and (ii) every KKT point of \(\mathcal H\) projects to a KKT point of \(P\).

We now outline the main steps of the reduction. In Section 4.1, we construct a normalized quadratic polynomial \(\widetilde P\) whose coefficients admit a decomposition into signed dyadic terms of the form \(s2^{-k}\), with \(s\in\{-1,1\}\) and \(k\) being polynomially bounded in the encoding length of \(P\). Lemma \ref{lemma:normalized polynomial} establishes that \(\widetilde P\) and \(P\) have exactly the same KKT points.

In Section 4.2, we introduce the quadratic chain gadget, which is the main technical tool of our reduction. For any dyadic monomial \(2^{-k}xy\), the gadget constructs a quadratic polynomial with constant-bounded integer coefficients and \(O(k)\) auxiliary variables. We prove that minimizing this polynomial over its auxiliary variables reproduces \(2^{-k}xy\) exactly, as established in Lemma \ref{lem: g equals G with minimizer}.

In Section 4.3, we combine the quadratic chain gadgets corresponding to all dyadic monomials of \(\widetilde P\). We show that every KKT point of the resulting quadratic polynomial projects onto a KKT point of \(\widetilde P\), and hence of \(P\). Finally, in Section 4.4, we transform the auxiliary domain to the unit hypercube and establish that the resulting polynomial \(\mathcal H\) has integer coefficients with polynomially bounded unary encoding lengths.

\subsection{Normalizing the Coefficients and Dyadic decomposition}\label{sec:normalized and dyadic coefficients}
Consider an instance of \qkkt\  and let \(P(\mathbf{x})\) be the corresponding quadratic polynomial over \([0,1]^n\), as defined in Equation (\ref{the original polynomial}). Let \(\operatorname{KKT}(P)\) denote the set of KKT points of \(P\). We construct a polynomial \(\widetilde P\) whose coefficients are polynomially bounded in the encoding length of \(P\), while preserving the set of KKT points, i.e.,
\(
\operatorname{KKT}(P)=\operatorname{KKT}(\widetilde P).
\)

Let \(A\) be the least common multiple of the denominators of the coefficients appearing in \(P(\mathbf x)\). Compute the polynomial \(P'(\mathbf x)  = A P(\mathbf x)\). Notice that \(P'\) admits only integer coefficients. Let \(M\) be the maximum value among 1 and the absolute values of the coefficients in \(P'\). Set \(L=\lfloor\log_2 M\rfloor+1\) and define \(S=2^L\).  Finally compute the polynomial \(\widetilde P(\mathbf x) = P'(\mathbf x)/S\). For this transformation the following holds. 

\begin{lemma}\label{lemma:normalized polynomial}
The polynomials \(P\) and \(\widetilde P\) have exactly the same KKT points, namely 
\(\operatorname{KKT(P)}=\operatorname{KKT}(\widetilde P)\). Moreover every coefficient \(d\) of \(\widetilde P\) admits a dyadic decomposition of the form 
\begin{equation}\label{dyadic decomposition}
d=s\sum_{k=1}^{L}\delta_k 2^{-k},
\end{equation}
where \(s\in\{-1,+1\}\) and \(\delta_k\in\{0,1\}\).
\end{lemma}

\begin{proof}
For every \(\mathbf x\in[0,1]^n\) and every \(i\in[n]\), we have
\(\frac{\partial\widetilde P}{\partial x_i}(\mathbf x) = \frac{A}{S} \frac{\partial P}{\partial x_i}(\mathbf x)\). Since \(A/S>0\), the corresponding partial derivatives of \(P\) and \(\widetilde P\) have the same sign and are equal to 0 at the same points. Consequently, for every \(i\in[n]\), the KKT condition associated with \(x_i\) is satisfied by \(P\) at \(\mathbf x\) if and only if it is satisfied by \(\widetilde P\) at \(\mathbf x\) and thus \(\operatorname{KKT(P)} = \operatorname{KKT(\widetilde P)}\). 

Since \(P'\) has integer coefficients and \(\widetilde P(\mathbf x)=P'(\mathbf x)/2^L\), every coefficient \(d\) of \(\widetilde P\) can be expressed as \(d = z/2^L\),
for some integer \(z\in\mathbb Z\) or which \(|z|<2^L\). If \(z=0\), the decomposition holds by choosing \(s=1\) and \(\delta_k=0\) for every \(k\in[L]\). Otherwise, let \(s\in\{-1,+1\}\) denote the sign of \(z\). Since \(0<|z|<2^L\), the binary representation of \(|z|\) contains at most \(L\) bits. Therefore, there exist \(\delta_1,\ldots,\delta_L\in\{0,1\}\) such that
\(|z|=\sum_{k=1}^{L}\delta_k2^{L-k} \). Dividing both sides by \(2^L\) and multiplying by \(s\), we get \( 
d= s\sum_{k=1}^{L}\delta_k2^{-k}\).
\end{proof}

Using the above lemma, we can express
$\widetilde{P}$ as a sum of monomials with signed dyadic coefficients. Let $R$ denote the total number
of nonzero terms obtained from expanding each coefficient in \(\widetilde P\) according to (\ref{dyadic decomposition}). For each $r\in[R]$, let
$p_r(\mathbf{x}),q_r(\mathbf{x})
\in\{1,x_1,\ldots,x_n\}$.
Their products represent all types of monomials appearing in \(\widetilde P\), namely, quadratic, linear, and constant monomials. Consequently, we can express \(\widetilde P\) as the following dyadic expansion.
\begin{equation}\label{eq:dyadic_expansion}
\widetilde{P}(\mathbf{x})
=\sum_{r=1}^{R}s_r2^{-k_r}
p_r(\mathbf{x})q_r(\mathbf{x}),
\qquad
s_r\in\{-1,+1\},\quad
1\leq k_r\leq L.
\end{equation}
Since each coefficient contributes at most \(L\) terms to the dyadic expansion and \(\widetilde P\) contains \(O(n^2)\) monomials, we have \(R=O(n^2L)\). Moreover, because \(L\) is polynomially bounded in the encoding length of \(P\), the dyadic expansion can be computed in polynomial time.

\subsection{Quadratic chain gadget}
Consider a dyadic monomial \(g(x,y) = s2^{-k}xy\), where \(s\in\{-1,1\}\), \(1\le k\leq L\) and \(x,y\in[0,1]\). Let \ \(m = k+2\) and introduce \(m\) auxiliary variables \(\mathbf u = (u_1,\ldots u_m)\in [-1,1]^m\). Consider the following quadratic polynomial which we refer to as \emph{quadratic chain} gadget. 

\begin{equation}\label{quadratic chain gadget}
G(x,y,\mathbf u) = 3 \sum_{i=1}^{m-1}(2u_i-u_{i+1})^2+9u_m^2-6x u_1+6sy u_{m}+x^2+y^2.
\end{equation}

For fixed values of \(x\) and \(y\) in \([0,1]\), let \(G_{x,y}:[-1,1]^m\to\mathbb R\) denote the quadratic chain gadget, viewed as a function of \(\mathbf u\). Notice that the quadratic polynomial $G_{x,y}$ is continuous over the compact set \([-1,1]^m\), and hence admits at least one global minimizer in this domain. Let \(\mathbf u^*(x,y)\) denote a global minimizer of \(G_{x,y}\), that is,
\begin{equation}\label{minimizer}
\mathbf u^*(x,y)\in
\operatorname*{arg\,min}_{\mathbf u\in[-1,1]^m}
G_{x,y}(\mathbf u).
\end{equation}

The next lemma shows the relation between a global minimizer and a KKT point of \(G_{x,y}\).

\begin{lemma}\label{minizer=kkt}
For every fixed \(x,y\in[0,1]\), the quadratic polynomial \(G_{x,y}\) admits a unique global minimizer \(\mathbf u^*(x,y)\) over \([-1,1]^m\), which is also the unique KKT point of \(G_{x,y}(\mathbf u)\) over \([-1,1]^m\).
\end{lemma}

\begin{proof}
First we will show that \(G_{x,y}\) is strictly convex in \([-1,1]^m\). Fix the values for the variables \(x,y\) and let \(Q(\mathbf u)\) denote the quadratic part of \(G_{x,y}\),  i.e.,

\[ 
Q(\mathbf u) = 3 \sum_{i=1}^{m-1}(2u_i-u_{i+1})^2+9u_m^2.
\]

Since \(G_{x,y}\) differs from \(Q\) only by linear terms in \(\mathbf u\), to establish that \(G_{x,y}\) is strictly convex, it suffices to show that \(Q\) is a positive-definite quadratic form, i.e., \(Q(\mathbf u)>0\) for every \(\mathbf u\neq\mathbf 0\)~\footnote{We refer the reader to Section 3.1.4, Example 3.2, page 71 in \cite{boyd2004convex}}.
Now observe that \(Q(\mathbf u)\geq0\) as a sum of square terms while the equality holds if and only if \(\mathbf u=\mathbf0\). Thus \(Q\) is a positive definite form and hence \(G_{x,y}\) is strictly convex. As a consequence of strict convexity, \(G_{x,y}\) admits at most one global minimizer in $[-1,1]^m$.~\footnote{see Section 4.2.1, page 137~\cite{boyd2004convex}.} Since \(G_{x,y}\) admits at least one global minimizer in \([-1,1]^m\), then \(\mathbf u^*(x,y)\) is the unique global minimizer of $G_{x,y}(\textbf{u})$, namely
\[\textbf{u}^*(x,y) = \arg \min_{\mathbf u \in[-1,1]^m}G_{x,y}(\mathbf u).\]

Finally since \(G_{x,y}\) is differentiable and strictly convex over \([-1,1]^m\), a point satisfies the KKT conditions for \(G_{x,y}\) if and only if it is a global minimizer\footnote{see Section 5.5.3 page 244 of \cite{boyd2004convex}}. As \(\mathbf u^*(x,y)\) is the unique global minimizer of \(G_{x,y}\), it follows that it is also the unique KKT point of \(G_{x,y}\).
\end{proof}

However, uniqueness on its own does not suffice to establish the correctness of the gadget. Since the auxiliary variables are box-constrained a KKT point of \(G\), with $x$ and $y$ not fixed, does not necessarily satisfy the stationarity equations of the original polynomial. We must therefore ensure that the auxiliary variables satisfy the {\em unconstrained stationarity} conditions, at every KKT of \(G\). 
This is satisfied if we prove that the unique minimizer \(\mathbf u^*\) for every fixed $x,y$ lies in the interior of the feasible domain. 

\begin{lemma}\label{lem:interior}
For every \(x,y\in[0,1]\) it holds that \(\mathbf u^*(x,y)\in (-1,1)^m\).
\end{lemma}
\begin{proof}
Fix \(x,y\in[0,1]\) and consider the unconstrained stationarity equations of \(G_{x,y}\) as follows
\begin{align}
4u_1-2u_2 &= x,
\label{eq:chain-left}\\
5u_i-2u_{i-1}-2u_{i+1} &= 0,
&& 2\le i\le m-1,
\label{eq:chain-interior}\\
4u_m-2u_{m-1} &= -sy.
&&
\label{eq:chain-right-attachment}
\end{align}

Since \(G_{x,y}\) is strictly convex, as established in Lemma \ref{minizer=kkt}, its unconstrained stationarity system \(\mathbf{LS}\) admits at most one solution. As a consequence if \(\mathbf{LS}\) admits a solution, then this is unique. 

We express the solution of \(\mathbf{LS}\) as a linear combination of two auxiliary solutions. Specifically  let \(\mathbf{LS1}\) denote the linear system obtained from \(\mathbf{LS}\) by setting \(x=1\) and \(y=0\) and let \(\mathbf{LS2}\) denote the linear system obtained from \(\mathbf{LS}\) by setting \(x=0\) and \(sy=-1\). Observe that all three linear systems share the same coefficient matrix, which is equal to the Hessian of \(Q\) multiplied by 1/6. Since \(Q\) is a strictly convex quadratic form, its Hessian, and as a consequence the coefficient matrix of the linear systems, is positive definite and hence invertible. Consequently, each system admits a unique solution.

Consider \(\mathbf v\) and \(\mathbf w\) to be the solutions of \(\mathbf{LS1}\) and \(\mathbf{LS2}\), respectively. By linearity we get that

\begin{equation}\label{u}
\mathbf u=x \mathbf v-sy\mathbf w.
\end{equation}

We now obtain the explicit solutions of \(\mathbf{LS1}\) and \(\mathbf{LS2}\). Observe that both systems share the following linear homogeneous recurrence relation
\[
5z_i-2z_{i-1}-2z_{i+1}=0,
\qquad 2\le i<m.
\]
where we use \(z_i\) to denote either \(v_i\) or \(w_i\). We solve this recurrence relation using the roots of its characteristic polynomial.
Assume that \(z_i = r^i\) for \(r^i\neq0\). Substituting \(z_i\) with \(r^i\) in the recurrence relations and dividing by \(r^{i-1}\) we get the characteristic quadratic polynomial 
\[2 r^2-5 r+2 = 0.\] 
Solving this polynomial we get the characteristic roots \(r = 1/2\) and \(r = 2\). Finally a solution to the homogeneous recurrence relation is given by the formula 
\begin{equation}\label{characteristic root}
z_i = a 2^i+b2^{-i},\quad a,b\in\mathbb Q ,\quad i\in[m].
\end{equation}
We proceed to determine the parameters \(a\) and \(b\) using the boundary conditions in \textbf{LS1} and \textbf{LS2}. Consider \(\mathbf{LS1}\) whose boundary conditions are
\(4v_1-2v_2=1\text{ and }4v_m-2v_{m-1}=0
\) and let \(u_i = a_12^i+b_12^{-i}\) for \(i\in[m]\) be a solution. Setting \(z_i\) as \(v_i\) and substituting (\ref{characteristic root}) in the first boundary condition  we get \(b_1=2/3\). Similarly substituting (\ref{characteristic root}) in the second boundary condition we get \(a_1=0\) and thus
\begin{equation}\label{v}
v_i=\frac{2^{1-i}}{3},
\qquad i\in[m].
\end{equation}
Next, consider \(\mathbf{LS2}\), whose boundary conditions are \(4w_1-2w_2=0\) and \(4w_m-2w_{m-1}=1\) and let \(w_i=a_2 2^i+b_2 2^{-i}\) for \(i\in[m]\) be a solution.  Setting \(z_i\) as \(w_i\) and substituting (\ref{characteristic root}) in the first boundary condition  we get \(b_2=0\). Similarly substituting (\ref{characteristic root}) in the second boundary condition we get \(a_2= 1/(32^{m})\) and therefore

\begin{equation}\label{w}
w_i=\frac{2^{i-m}}{3},
\qquad i\in[m].
\end{equation}
Substituting (\ref{v}) and (\ref{w}) in (\ref{u}) we get that the unique solution of \textbf{LS} is 
\begin{equation}\label{explicit u}
u_i=\frac13\Big(2^{1-i} x-s2^{i-m}y\Big),\quad i\in[m].
\end{equation}
Finally since \(x,y\in[0,1]\), applying the triangle inequality for  every \(i\in[m]\),
\[
|u_i| \leq \frac{1}{3}\big(2^{1-i}+2^{i-m}\big)\leq \frac23,
\]
which proves that \(\mathbf u ^*(x,y)\in (-1,1)^m\).
\end{proof}

Having established the interiority of the unique minimizer \(\mathbf u^*(x,y)\), we proceed to show that substituting \(\mathbf u^*(x,y)\) into the quadratic chain gadget \(G(x,y,\mathbf u)\) yields exactly the original dyadic monomial \(g(x,y)\).

\begin{lemma}\label{lem: g equals G with minimizer} For every  \(x\) and \(y\) in \([0,1]\), it holds that \(G(x,y,\mathbf u^*(x,y)) =  g(x,y)\).
\end{lemma}

\begin{proof}
We substitute the explicit form of \(\mathbf u^*(x,y)\) as given in (\ref{explicit u}), in the quadratic chain gadget expression (\ref{quadratic chain gadget}). For \(1\le i<m\) it holds that 
\[
2 u_i^*-u^*_{i+1} =\frac13\left(2^{2-i} x-s2^{i-m+1} y\right)-\frac13 \left(2^{-i} x-s2^{i-m+1} y\right)=2^{-i} x.
\]
Therefore,
\[
3 \sum_{i=1}^{m-1}(2u_i^*-u_{i+1}^*)^2=3x^2 \sum_{i=1}^{m-1}4^{-i}=x^2 (1-4^{1-m}).
\]
Adding the explicit expression \(u^*_m = \frac{1}{3} (2^{1-m} x-sy)\) and \(u^*_1 = \frac{1}{3} (x-s2^{1-m} y )\) we get that 
\[
G(x,y,\mathbf u^*(x,y)) = x^2 (1-4^{1-m})
+(2^{1-m} x-sy)^2-2x(x-s2^{1-m}y)+2sy (2^{1-m} x -sy)+x^2+y^2,
\]
since \(m = k+2\) after calculations we get that 
\(
G(x,y,\mathbf u^*(x,y)) = s2^{-k} x y.
\)
\end{proof}

\subsection{Computing a KKT point of \(\widetilde{P}\)}
Recall that the dyadic decomposition of \(\widetilde P(\mathbf x)\), given in (\ref{eq:dyadic_expansion}), consists of \(R\) monomials of the form \(g_r\big(p_r(\mathbf x),q_r(\mathbf x)\big) = s_r2^{-k_r}p_r(\mathbf x)q_r(\mathbf x)\), with \(r\in[R]\) and \(1\le k_r\leq L \). For each monomial \(g_r\) consider a set of \(m_r = k_r+2\) disjoint auxiliary variables  \(\mathbf{u}_r = \big(u_{r,1},\ldots u_{r,m_r}\big)\in[-1,1]^{m_r}\) and construct the quadratic chain gadget \(G_r\bigl(p_r(\mathbf x),q_r(\mathbf x),\mathbf u_r\bigr)\) as given in (\ref{quadratic chain gadget}). Letting \(\mathbf u=(\mathbf u_1,\ldots,\mathbf u_R)\) and \(\mathcal U  = \prod_{r=1}^{R}[-1,1]^{m_r}\), we add the quadratic chain gadgets and get the induced \emph{combined quadratic chain polynomial}

\begin{equation}\label{combined quadratic chain polynomial }
\mathcal G(\mathbf x,\mathbf u) = \sum_{r=1}^{R}
G_r\bigl(p_r(\mathbf x),q_r(\mathbf x),\mathbf u_r\bigr),\quad \text{ where }\quad  (\mathbf x,\mathbf u)\in [0,1]^n \times \mathcal U
\end{equation}

Notice that because each quadratic chain gadget admits a disjoint set of auxiliary variables,  then for fixed \(\mathbf x\), minimizing \(\mathcal G(\mathbf x,\mathbf u)\) over \(\mathbf u\) is equivalent to minimizing each \(G_r\) over \(\mathbf u_r\). Since for fixed \(\mathbf x\in[0,1]^n\) by the notation we defined earlier
\[
\min_{\mathbf u_r\in[-1,1]^{m_r}}G_r(p_r(\mathbf x),q_r(\mathbf x),\mathbf u_r) = G_r\Big(p_r(\mathbf x),q_r(\mathbf x),\mathbf{u}_r^*\big(p_r(\mathbf x),q_r(\mathbf x)\big)\Big),
\]
then combining the above expression with Lemma \ref{lem: g equals G with minimizer} we get that

\begin{equation}\label{G_r and g_r}
\sum_{r=1}^{R}
\min_{\mathbf u_r\in[-1,1]^{m_r}}
G_r(p_r(\mathbf x),q_r(\mathbf x),\mathbf u_r)=
\sum_{r=1}^{R}
g_r(p_r(\mathbf x),q_r(\mathbf x)).
\end{equation}

For any \(\mathbf x\in[0,1]^n\), let \(\mathbf u^*(\mathbf x) = \big(\mathbf u^*_1(\mathbf x),\ldots ,\mathbf {u}^*_R(\mathbf x)\big)\), where \(\mathbf u_r^*(\mathbf x)\) denotes the unique auxiliary minimizer of the \(r\)-th quadratic chain gadget. Then for a fixed vector \(\mathbf x\in[0,1]^n\), 
\begin{equation*}
\mathcal G(\mathbf x,\mathbf u^*(\mathbf x)) = \min_{\mathbf u\in\mathcal U}\mathcal G(\mathbf x,\mathbf u) = \sum^R_{r=1}\min_{\mathbf u_r\in[-1,1]^{m_r}}G_r(p_r(\mathbf x),q_r(\mathbf x),\mathbf u_r),  
\end{equation*}
where from equations (\ref{G_r and g_r}) and (\ref{eq:dyadic_expansion}) we get that 
\(
\mathcal G\big(\mathbf x,\mathbf u^*(\mathbf x)\big) = \widetilde P(\mathbf x)
\). Finally we establish that every KKT point of \(\mathcal G\) maps back to an exact KKT point of \(\widetilde P\).

\begin{proposition}\label{prop:KKT approximation}
If \((\mathbf x',\mathbf u')\) is a KKT point of \(\mathcal G\) then \(\mathbf x'\) is a KKT point of \(\widetilde P\).
\end{proposition}

\begin{proof}
Let \((\mathbf x',\mathbf u')\) be a KKT point of \(\mathcal G\). Since the auxiliary variables of the quadratic chain gadgets are disjoint, for every \(r\in[R]\), the vector \(\mathbf u_r'\) satisfies the KKT conditions of \(G_r(p_r(\mathbf x'),q_r(\mathbf x'),\mathbf u_r)\) with respect to \(\mathbf u_r\). From Lemma \ref{minizer=kkt}, each quadratic chain gadget admits a unique KKT point with respect to its auxiliary variables which coincides with its unique interior minimizer. Consequently,
\(\mathbf u_r'=\mathbf u_r^*(\mathbf x')\text{ for every }r\in[R],\) and hence \(\mathbf u'=\mathbf u^*(\mathbf x')\). From the previous discussion it holds that for every \(\mathbf x\in[0,1]^n\),
\(\widetilde P(\mathbf x) = \mathcal G(\mathbf x,\mathbf u^*(\mathbf x)).
\) Differentiating with respect to \(x_i\), for \(i\in[n]\), we get that 
\begin{equation*}
\frac{\partial\widetilde P}{\partial x_i}(\mathbf x')
=
\frac{\partial\mathcal G}{\partial x_i}
(\mathbf x',\mathbf u^*(\mathbf x'))+
\sum_{r=1}^{R}\sum_{j=1}^{m_r}
\frac{\partial\mathcal G}{\partial u_{r,j}}
(\mathbf x',\mathbf u^*(\mathbf x'))
\frac{\partial u_{r,j}^*}{\partial x_i}(\mathbf x').
\end{equation*}

Because in Lemma \(\ref{lem:interior}\), we showed that the auxiliary minimizers are interior therefore
\(
\frac{\partial\mathcal G}{\partial u_{r,j}}
(\mathbf x',\mathbf u^*(\mathbf x'))=0
\) for every \(r\in[R]\) and \(j\in[m_r]\). Hence for every \(i\in[n]\) it follows that
\begin{equation}\label{derivative of athcal G is drivatie of widetilde P}
\frac{\partial\widetilde P}{\partial x_i}(\mathbf x')
=
\frac{\partial\mathcal G}{\partial x_i}
(\mathbf x',\mathbf u').\quad 
\end{equation}
Since \((\mathbf x',\mathbf u')\) is a KKT point of \(\mathcal G\), it follows from Equation (\ref{derivative of athcal G is drivatie of widetilde P}) that \(\mathbf x'\) satisfies the KKT conditions of \(\widetilde P\) over \([0,1]^n\).
\end{proof}

\subsection{Transforming the auxiliary domain}

To finalize the reduction we transform the domain of the auxiliary variables from \([-1,1]^m\) to \([0,1]^m\).
Let \(M=\sum_{r=1}^{R}m_r\) denote the total number of auxiliary variables in \(\mathcal G(\mathbf x,\mathbf u)\). For each auxiliary variable \(u_{r,j}\in [-1,1]\), we introduce a new variable \(v_{r,j}\in[0,1]\) such that \(u_{r,j}=2 v_{r,j}-1\). Let \(\mathbf v\) denote the vector of all new auxiliary variables and let 
\(
{\mathcal H}(\mathbf x,\mathbf v)
=
\mathcal G(\mathbf x,2\mathbf v-\mathbf 1)
\) be the resulting quadratic polynomial defined over \([0,1]^{n+M}\). For every \(i\in[n]\) and every auxiliary coordinate \((r,j)\),
\[
\frac{\partial{\mathcal H}}{\partial x_i}
(\mathbf x,\mathbf v)
=
\frac{\partial\mathcal G}{\partial x_i}
(\mathbf x,2\mathbf v-\mathbf 1)
\quad\text{ and }\quad
\frac{\partial{\mathcal H}}{\partial v_{r,j}}
(\mathbf x,\mathbf v)
=
2\frac{\partial\mathcal G}{\partial u_{r,j}}
(\mathbf x,2\mathbf v-\mathbf 1).
\]
Also the substitution maps the boundary points of the two intervals as follows:
\(
v_{r,j}=0\) if and only if \(u_{r,j}=-1\) and \(v_{r,j}=1\) if and only if \(u_{r,j}=1\), while every interior point is mapped again to an interior point. Consequently, the KKT conditions for the original variables remain unchanged, while the KKT conditions for the auxiliary variables are preserved because their partial derivatives are multiplied by \(2\). Hence,
\begin{equation}\label{final KKT correspondence}
(\mathbf x,\mathbf v)\in
\operatorname{KKT}({\mathcal H})
\iff
(\mathbf x,2\mathbf v-\mathbf 1)\in
\operatorname{KKT}(\mathcal G).
\end{equation}
\subsection{Proof of Theorem \ref{main theorem}}

Consider an arbitrary instance of \qkkt\ defined by a quadratic polynomial \(P\) over \([0,1]^n\), and let \(N\) denote its binary encoding length. By Lemma \ref{lemma:normalized polynomial}, the polynomial \(\widetilde P\) constructed in Section 4.1 has exactly the same KKT points as \(P\) and admits a dyadic decomposition consisting of \(R=O(n^2L)\) monomials. 

We first show that \(L=O(N)\). Recall from Section 4.1 that \(A=\operatorname{lcm}(b_1,\ldots,b_t)\), where let's assume that \(b_1,\ldots,b_t\) are the denominators of the coefficients of \(P\), and that \(M\) is the maximum of 1 and the absolute values of the coefficients of \(P'=A P\), while \(L=\lfloor\log_2 M\rfloor+1\). Since \(A\le\prod_{j=1}^{t}b_j\), we have \(\log_2 A\le\sum_{j=1}^{t}\log_2 b_j\le N\). As the numerators of \(P\) also have binary encoding lengths at most \(N\), every coefficient of \(P'=A P\) has absolute value at most \(2^{O(N)}\). Consequently, \(\log_2 M=O(N)\), and hence \(L=O(N)\).

Next, we construct the combined quadratic chain polynomial \(\mathcal G\) as defined in Equation (\ref{combined quadratic chain polynomial }), that is summing over all quadratic chain gadgets~(see Equation (\ref{quadratic chain gadget})) that correspond to the dyadic monomials of \(\widetilde P\). Let \(M_{aux}\) denote the total number of auxiliary variables that emerge from the construction of \(\mathcal G\). By Proposition \ref{prop:KKT approximation}, every KKT point of \(\mathcal G\) projects onto a KKT point of \(\widetilde P\), and hence, by Lemma \ref{lemma:normalized polynomial}, to a KKT of \(P\). Applying the affine transformation of Section 4.4, we obtain the quadratic polynomial \(\mathcal H(\mathbf x,\mathbf v)
=\mathcal G(\mathbf x,2\mathbf v-\mathbf 1)\) over the domain \([0,1]^{n+M_{aux}}\). By Equation (\ref{final KKT correspondence}), every KKT point of \(\mathcal H\) corresponds to a KKT point of \(\mathcal G\), and therefore yields a KKT point of \(P\).

It remains to bound the encoding length of the constructed instance \(\mathcal H\). Since each gadget introduces \(m_r=k_r+2\le L+2\) auxiliary variables, their total number satisfies
\[
M_{\mathrm{aux}}
=\sum_{r=1}^{R}m_r
\le R (L+2)
=O(n^2L^2).
\]

Moreover, each gadget contains \(O(m_r)\) terms with constant-bounded integer coefficients~(see Equation (\ref{quadratic chain gadget})), which also holds after the affine transformation. Thus, every coefficient of the resulting polynomial \(\mathcal H\) is an integer of absolute value at most
\[
O\left(\sum_{r=1}^{R}m_r\right)
=O\big(n^2L^2\big).
\]

Since we established that \(L=O(N)\), both the number of variables and the unary encoding lengths of the coefficients of \(\mathcal H\) are polynomially bounded in \(N\), i.e, the encoding length of \(P\). Since \(\mathcal H\) is quadratic and has polynomially many variables, then it contains at most quadratically many monomials in the number of variables. Furthermore, its construction can be performed in polynomial time, which concludes our proof.
 
\section{Discussion}
One obvious and important question that remains open is whether \qkkt~ remains hard, either for $\CLS$, or for a different complexity class/under a different assumption, when the coefficients are {\em constant}. 
Such a result would not contradict anything of our current understanding of $\CLS$, or $\TFNP$ in general.
As a side note, let us highlight two things. Firstly, quadratic {\em optimization} is $\NP$-hard even when all coefficients are constants (think of Max-Cut in unweighted graphs). Second, there exist several problems that are $\PPAD$-hard even when all coefficients are small constants~\cite{deligkas2024pure} or even binary~\cite{PapadimitriouP21-threshold-games,DinhH24,batziou2026complexity}.

One approach to attack the above mentioned problem would be to use our technique when the starting point is a quadratic polynomial with polynomially-bounded coefficients and hope that after a number of iterations we can end up with a polynomial with constant coefficients. 
Unfortunately, this cannot work under the current construction of the chain gadget. 
The reason is that after the dyadic decomposition, for every occurrence of a variable $x$ it creates one copy of $x^2$ in every chain. 
One way to bypass this obstacle could be by proving that \qkkt~ remains hard even on ``sparse'' instances, i.e., instances where each variable appears only a constant number of times.
Another direction could be to create a different chain gadget that does not preserve a one-to-one correspondence of stationary points, but works only the approximate sense.
 
\section*{Acknowledgements}
We sincerely thank Ioannis Panageas for the insightful discussions on $\CLS$ and for his substantial contribution to the development of the final form of the paper.
Argyrios Deligkas and Stavros D. Ioannidis were supported by the EPSRC grant EP/X039862/1. 
We used various versions of GPT Pro to explore proof strategies during the development of this work.

\bibliographystyle{alpha}
\bibliography{bibl}
\end{document}